\documentclass[preprint]{ptephy_v1}

\preprintnumber{XXXX-XXXX} 
\usepackage{hyperref}

\begin{document}
\title{Bifurcation and anomalous spectral accumulation in oval billiard.}
%
%
\author{Hironori Makino}
\affil{Department of Human and Information Science at Tokai University\email{makino@tokai-u.jp}}
%
%
%

\begin{abstract}
Spectral statistics of quantum oval billiard whose classical dynamical system shows bifurcations is numerically investigated in terms of the two-point correlation function (TPCF) which is defined as the probability density of finding two levels at a specific energy interval. The eigenenergy levels at bifurcation point is found to show anomalous accumulation which is observed as a periodic spike oscillation of the TPCF. We analyzed the eigenfunctions localizing onto the various classical trajectories in the phase space and found that the oscillation is supplied from a limited region in the phase space, which contains the bifurcating orbit. We also show that the period of the oscillation is in good agreement with the period of a contribution from the bifurcating orbit to the semiclassical TPCF obtained by Gutzwiller trace formula. 
\end{abstract}
\subjectindex{xxxx, xxx}
\maketitle
\section{Introduction}
It has been conjectured that in the semiclassical limit, eigenenergy levels of quantum system whose classical dynamical system is integrable, are equivalent to the uncorrelated random numbers from the Poisson process and are well characterized by the Poisson statistics\cite{BT}, while the eigenenergy levels of classically fully-chaotic system repel each other and are well characterized by the Gaussian-orthogonal-ensemble(GOE) statistics or Gaussian-unitary-ensemble(GUE) statistics of random matrix theory\cite{BGS}. In between these two extremes is a generic case of mixed type dynamics where regular and chaotic orbits coexist in the classical phase space. In this case, it has been proposed that the quantum level statistics is a combination of the Poisson and GOE/GUE statistics\cite{BR,SV}, with relative weights determined by the corresponding phase volumes (Liouville measures) of the regular and chaotic orbits\cite{Robnik1998}. Our concern in this paper is to identify in this mixed case anomalous accumulation of levels that is associated with periodic orbit bifurcations.

Bifurcation is a characteristic phenomenon of the mixed dynamical systems where periodic orbits are created or destroyed by coalescence, and this process provides a very important perspective to the Gutzwiller semiclassical periodic-orbit theory which connects the fluctuation properties of the quantum energy spectrum to the periodic-orbit sum\cite{Gutz}. In several investigations\cite{Scho,Scho2,BKP,Makino1999}, it has been revealed that the semiclassical contribution from the bifurcating orbit to the quantum level statistics becomes relatively large in the semiclassical limit, and the periodic-orbit generation through the bifurcation has a large impact on the quantum fluctuation statistics. Berry, Keating and Prado investigated for the first time, the effect of bifurcation on the energy level statistics\cite{BKP}. They analyzed the level number variance(LNV) of the perturbed cat map at a saddle-node (tangent) bifurcation and found that there are additional contributions to the long-range spectral correlation called {\it lift-off} effect which came from the bifurcating periodic orbit.

The quantum mechanical effects of bifurcation have also been reported in the study of short-range spectral correlations. Makino, Harayama and Aizawa investigated numerically the nearest-neighbor level-spacing distribution (NNLSD) in quantum oval billiard and found anomalous accumulation of adjacent levels at the bifurcation point where the hyperbolic periodic orbit is created\cite{Makino1999}.  However, its mechanism leading to the accumulation  was not well understood in terms of the semiclassical theory, since the NNLSD they focused on, cannot be described directly using the periodic orbit sum.  In this regard, it is better to focus on other statistical observables instead, such as two-point correlation function(TPCF), LNV and spectral rigidity\cite{Mehta}.

In this paper by the numerical experiments for the quantum oval billiard, we will analyze the TPCF whose semiclassical theory is well established in terms of the periodic orbit theory and show that the periodic orbit generation at the bifurcation causes remarkable spike-oscillations, which correspond to the spectral accumulation of levels reported by Makino et al. in the studies of NNLSD\cite{Makino1999,Makino2018}, and also to the {\it lift-off} effect of the long-range correlation reported in the study of the LNV by Berry, Keating and Prado\cite{BKP}.

This paper is organized as follows. In Sect.2, the classical dynamical system of the oval billiard is introduced, where the Poincar\'e surface of section and the bifurcation are analyzed.  The statistical property of eigenenergy levels in quantum oval billiard is precisely analyzed in Sect.3, where the TPCF shows remarkable oscillations at the bifurcation point.   We also discuss the relationship between the results of this paper and other related researches which analyzed the LNV and the NNLSD.  Finally, we conclude our work in Sect.4

\section{Bifurcation of classical dynamical system}
\begin{figure}[ttbp]
\begin{center}
\includegraphics{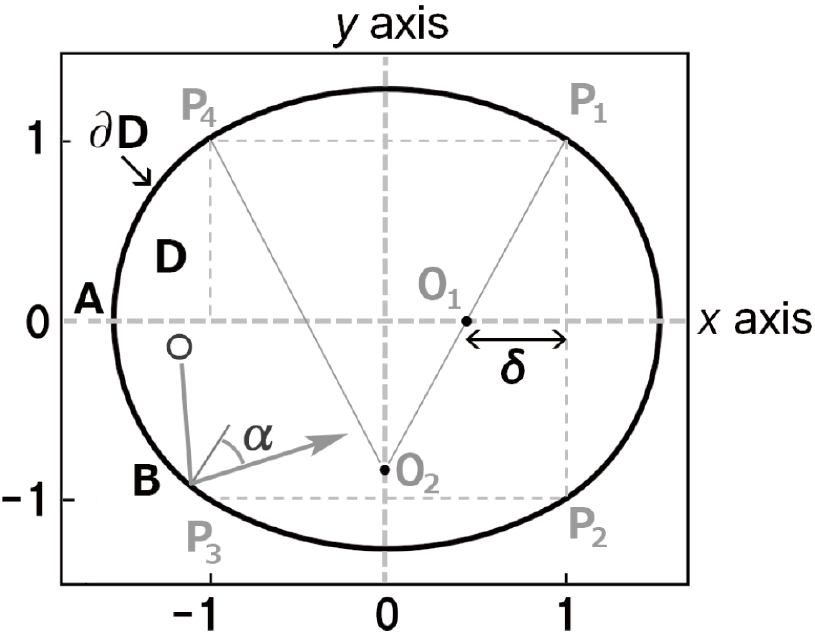}
\end{center}
\caption{Schematic definition of the billiard table D whose boundary wall $\partial$D consists of four circular arcs, $\stackrel{\textstyle\frown}{\mathrm{P_1 P_2}}$, $\stackrel{\textstyle\frown}{\mathrm{P_2 P_3}}$, $\stackrel{\textstyle\frown}{\mathrm{P_3 P_4}}$ and $\stackrel{\textstyle\frown}{\mathrm{P_4 P_1}}$. These arcs are connected smoothly at the four vertices P$_i, i=1-4$. $\delta\in[0,1]$ is a deformation parameter which determine the shape of $\partial$D.}
\label{fig1}
\end{figure}

The billiard table D is defined as follows(see also Figure \ref{fig1}): In the $xy$ plane, we consider a square whose vertices $\mbox{P}_i$, $i=1-4$ are located at $(\pm 1,\pm1)$. Let the point $\mbox{O}_1$ at $(1-\delta,0)$, $\delta\in[0,1]$, be the center of an arc $\stackrel{\textstyle\frown}{\mathrm{P_1 P_2}}$, and $\mbox{O}_2$ which is an intersection of the extension line of $\mbox{P}_1\mbox{O}_1$ and the $y$ axis, be the center of an arc $\stackrel{\textstyle\frown}{\mathrm{P_4 P_1}}$. The billiard wall $\partial$D is defined by the above two arcs and another two arcs, $\stackrel{\textstyle\frown} {\mathrm{P_3 P_4}}$ and $\stackrel{\textstyle\frown}{\mathrm{P_2 P_3}}$, constructed similarly. For $\delta=0$ where $\partial$D is a stadium shape, the system is strongly chaotic and the motion of a particle on D is unstable for any choice of initial condition\cite{Bunimovich}. On the other hand, for $\delta=1$ where $\partial$D is circular, the system is integrable and the motion of a particle on D is stable and regular for any choice of initial condition. For $0<\delta<1$, the billiard system is generic mixed type and the motion of a particle is regular or chaotic depending on the initial condition\cite{Benettin,Henon}. 

\begin{figure}[ttbp]
\begin{center}
\includegraphics{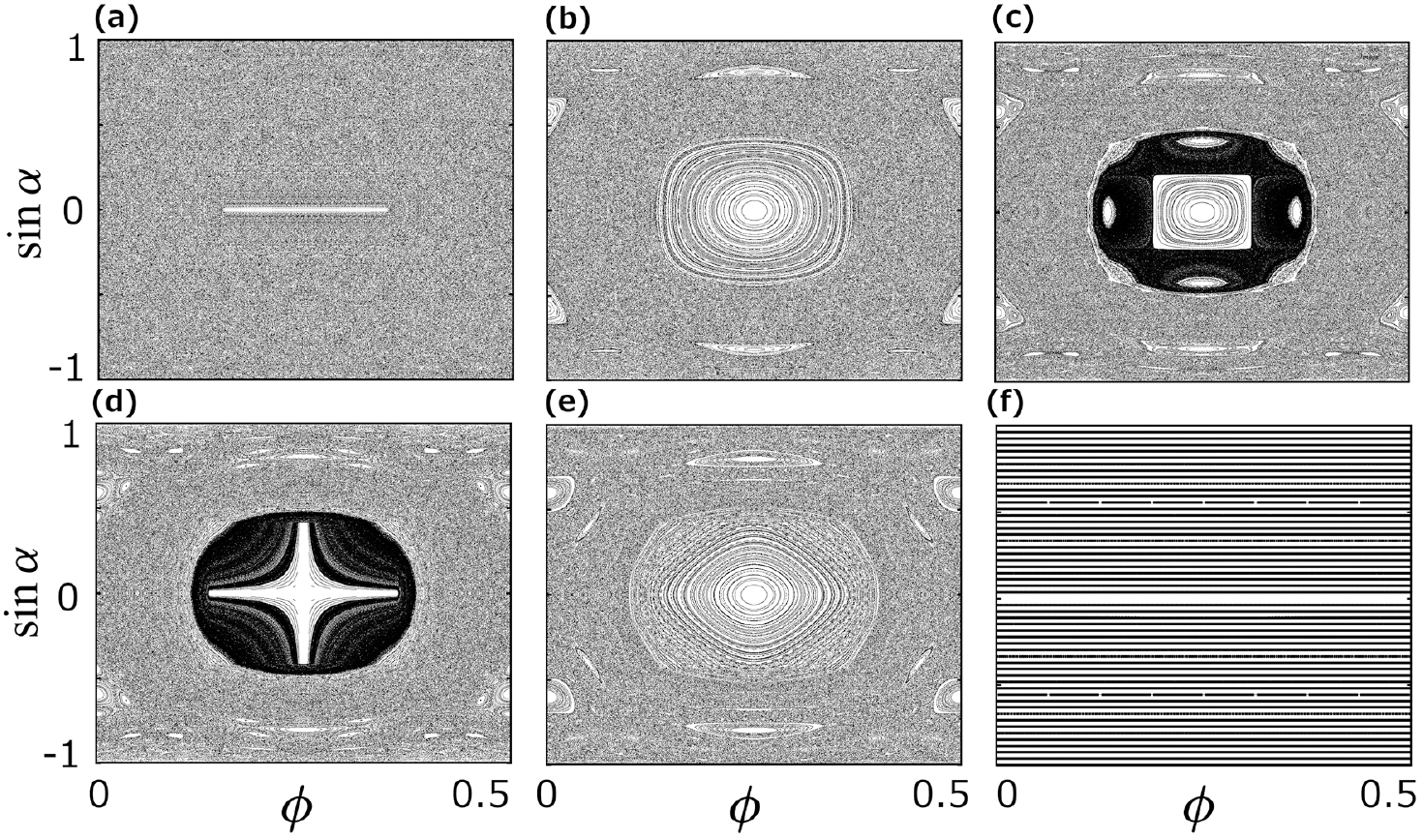}
\end{center}
\caption{Poincar\'e surface of section for (a) $\delta=0$, (b)$\delta=0.39$, (c)$\delta=0.445$, (d)$\delta=0.451\sim\delta_c$, (e)$\delta=0.475$, (f)$\delta=1$. }
\label{fig2}
\end{figure}

First we analyze the Poincar\'e surface of section(PSS) of the classical dynamical system. For 2D billiard problems, PSS is described by the Birkhoff coordinates $(\phi , \sin\alpha)$\cite{Birkhoff}, where $\phi\in[0,1]$ is the normalized curvilinear distance along the wall $\partial$D measured from the origin A to the collisional point B [see Fig.\ref{fig1}(a)], and $\alpha$ is the angle between the inner normal and the orbit reflected from the wall. Figures \ref{fig2}(a)-(f) are the trajectories in the Birkhoff coordinates for various values of $\delta$. In the cases $\delta=0$ (stadium shape) and $\delta=1$(circular shape), the entire surfaces of section are filled with chaotic and regular orbits, respectively[see Figs.\ref{fig2}(a) and \ref{fig2}(f)]. As $\delta$ increases from 0, one can observe the gradual enhancement of the island region through successive bifurcations[Fig.\ref{fig2}(b)]. However, this region suddenly shrinks at the critical point $\delta_c=(4-\sqrt{7})/3\sim 0.451$ as shown in Figs.\ref{fig2}(c) and \ref{fig2}(d), where the hyperbolic fixed points with period $4$ at
\begin{equation}
\phi_4^{\mp\pm}(\delta)=\frac{1}{2}\mp\frac{1}{4}+\frac{\alpha_4^{\pm}}{2\pi+4(\delta-1)\arctan{(1/\delta)}} ,\quad \sin\alpha_4^{\pm}(\delta) 
=\pm\frac{\sqrt{ 3\delta^2-8\delta+3}}{2(1-\delta)},\label{eq3}
\end{equation}
coalesce with the elliptic periodic points with period $2$ at $(\phi_2^{\pm}=0.5\pm0.25, \sin\alpha_2= 0) $. This bifurcation is very interesting since the remaining outer boundary of the island region generates the second chaotic component around the elliptic periodic points $(\phi_2^{\pm}, \sin\alpha_2) $. The bifurcation parameter $\delta_c$ has been determined by $\sin\alpha_4^{\pm}(\delta) =0$ or also by using the monodromy matrix $M(\delta)$ and the condition $|\mbox{tr}{M(\delta)}|=2$( see also Refs.\cite{Reichel,Makino2001}). When $\delta$ exceeds $\delta_c$, the hyperbolic fixed points $(\phi_4^{\mp\pm},\sin\alpha_4^{\pm})$ disappear into the imaginary domain beyond $\phi=\phi_2^{\pm}$ as shown in Fig.\ref{fig2}(e), and the island region is created again.

\section{Effect of bifurcation on energy level statistics}
Next, the quantum mechanical effect of the bifurcation is analyzed for the statistical property of energy levels, characterized by the TPCF. The eigenenergy levels $E_n, n=1,2,3\cdots$ are obtained by solving the time-independent Schr\"odinger-Helmholz equation $\nabla^2\psi({\bf r})+E\psi({\bf r})=0$ under the Dirichlet boundary condition $\psi({\bf r}\in\partial D)=0$, and they are transformed to a stationary point process $\{ \epsilon_n\}$ called {\it unfolded} energy levels whose mean spacing is unity\cite{Bohigas2}. The transformation $\{E_n\} \to \{ \epsilon_n\}$ is carried out by using the leading Weyl term of the integrated density of states, $N(E) ={\cal A}E/16\pi$, as $\epsilon_n = N(E_n)$, and for the quantum billiard problems, this transformation is equivalent to determining the area of a billiard table to $16\pi$. The TPCF is the statistical observable that stands for the probability density of finding two levels at spacing $L$, and is defined by using the density of states $d(\epsilon)=\sum_n\delta(\epsilon-\epsilon_n)$ as $R_2(L)=\left< d(\epsilon-L/2)d(\epsilon+L/2) \right>$, where the bracket $\left<\cdots\right>$ denotes an averaging over $\epsilon$. This quantity is useful to analyze the quantum mechanical effect of bifurcation since its relationship to the classical periodic orbits is well developed in the semiclassical theory. Based on the Gutzwiller's trace formula\cite{Gutz}, the TPCF is expressed by the periodic orbits sum as
\begin{eqnarray}
\left< d\left(\epsilon-\frac{L}{2}\right)d\left(\epsilon+\frac{L}{2}\right) \right>&\simeq & \sum_j 
| C_{j}(\epsilon) |^ 2 \cos{\left[\frac{T_j}{\hbar}L \right] }\nonumber\\
&& + O\left(\sum_{j_1\not=j_2}\left<\exp{\left[i\frac{S_{j_1}(\epsilon)-S_{j_2}(\epsilon) }{\hbar}\right] }\right>\right),
\label{eq:0}
\end{eqnarray}
where $j$ labels each primitive periodic orbit and its repetitions, $C_j (\epsilon)$ represents the amplitude associated to the monodromy matrix, $S_j(\epsilon) $ is the action integral along the orbit $j$, defined here to include the Maslov index, and $T_j =\partial S_j(\epsilon)/\partial\epsilon$ is the time period of the periodic orbit. Since the second term in the RHS of Eq.(\ref{eq:0}) is expected to vanish by the smoothing procedure over $\epsilon$, the TPCF is approximated by a sum of periodic functions $| C_{j}(\epsilon) |^ 2 \cos{\left[T_j L /\hbar \right] }$ whose periods with respect to $L$ are described as $p_j=2\pi\hbar/T_j$. These periods are rewritten for the billiard problem using the orbit length $l_j$ as $p_j=4\pi\sqrt{\epsilon}/l_j$. 

The creation of a new periodic orbit $j^*$ across the bifurcation generates an additional contribution $|C_{j^*}(\epsilon)|^2 \cos(T_{j^*}L/\hbar) $ to the RHS of Eq.(\ref{eq:0}), and this term may provide a large influence on the property of the TPCF if the amplitude $C_{j^*}(\epsilon)$ is not negligible in the infinite series. Such a possibility can arise at the bifurcation point where $C_{j^*}(\epsilon)$ becomes semiclassically large(see results of Refs.\cite{Scho,Scho2,BKP}).

Figures 3(a)-(c) show numerical plots of $R_2(L)$ for (a)$\delta=0$, (b)$0.39$ and (c)$1$, which are computed by the boundary integral method. the TPCF analyzed in this paper is finally determined by the superpositions of four TPCFs which are computed respectively from the {\it unfolded} energy levels obeying the four parity-symmetry-classes:$\psi(\pm x,y)=\pm\psi(x,y)$ and $\psi(x,\pm y)=\pm\psi(x,y)$. When the classical dynamical system is the strongly chaotic($\delta=0$), the TPCF is well approximated by $R_2^{\mbox{\tiny GOE}}(L)=1-\sigma^2(L)-\frac{d\sigma(L)}{dL}\int_L^{+\infty}\sigma(L')dL' $ with $\sigma(L)=\sin{(\pi L)}/(\pi L)$ which results from the GOE statistics[Fig.3(a)], while in the case of the integrable ($\delta=0$), the TPCF is well approximated by $R_2^{\mbox{\tiny Poisson}}(L)=1$[Fig.3(c)], which results from the Poisson statistics\cite{BT,BGS,Mehta}. For the mixed system($0<\delta<1$) whose classical phase space consists of regular and chaotic regions, the TPCF fits neither with $R_2^{\mbox{\tiny GOE}}(L)$ nor $R_2^{\mbox{\tiny Poisson}}(L)$. For this case, it is proposed that the eigenenergy spectrum is a combination of the Poisson and the GOE/GUE statistics whose TPCF is described by an interpolation formula\cite{Makino2009}: $R_2(\rho,L) =\rho R_2^{\mbox{\tiny GOE}}(\rho L) + (1-\rho) R_2^{\mbox{\tiny Poisson}}( (1-\rho) L) $, where the weight $\rho\in[0,1]$ is assumed to coincide with the relative phase volume(Liouvill\'e measure) of the chaotic component in the classical dynamical system having a single chaotic region\cite{Robnik1998}. In this paper we do not go into its physical meaning and deal with $\rho$ as a parameter. 

Figure 4 shows numerical plots of $\rho^{\mbox{\tiny qm}}(\delta)$, which are obtained by the best fitting curve of the formula $R_2(\rho,L)$ to the numerical data of TPCF. In each plot, we have used $6000$ levels from $\epsilon=8000$. The red curve is the relative Liouville measure $\rho^{\mbox{\tiny cl}}(\delta)$ of the chaotic region in the classical dynamical system, which is computed by the method introduced in Ref.\cite{Makino2018}. In the neighborhood of $\delta=0$, $\rho^{\mbox{\tiny qm}}(\delta)$ well agrees with $\rho^{\mbox{\tiny cl}}(\delta)$, whereas it largely deviates near the bifurcation point $\delta_c$ where the second chaotic region is generated in the classical dynamical system. The discrepancy between $\rho^{\mbox{\tiny qm}}(\delta)$ and $\rho^{\mbox{\tiny cl}}(\delta)$ is not surprising in itself since the formula $R_2(\rho,L)$ is designed for dynamical systems with a single chaotic component. The point of interest here is the sharp decrease of $\rho^{\mbox{\tiny qm}}(\delta)$ at $\delta_c$. This phenomenon was first reported in the study of the short-range spectral fluctuation characterized by the NNLSD, where the accumulation of levels between adjacent levels is observed\cite{Makino1999}. In the present paper, we will investigate this phenomenon more precisely using the TPCF.

Figures 5(a)-(c) show numerical plots of the TPCF $R_2(L)$ at the bifurcation point $\delta_c$, which are obtained for three different energy ranges; 6000 levels from $\epsilon=2000$ in Fig.5(a), 6000 levels from $\epsilon=8000$ in Fig.5(b) and 12000 levels from $\epsilon=4\times 10^5$ in Fig.5(c). The inset in each figure shows the enlargement around $R_2(L)=1$. The TPCF shows very interesting oscillation which generates the accumulation $R_2(L)>1$ of levels at a specific energy interval, and does not fit either $R_2^{\mbox{\tiny Poisson}}(L)$, $R_2^{\mbox{\tiny GOE}}(L)$ or their interpolation $R_2(\rho^{\mbox{\tiny qm}}, L)$ at all. In all energy ranges, the period of the oscillation is well approximated by the fundamental period $p=4\pi\sqrt{\epsilon}/l\sim0.46\sqrt{\epsilon}$ of the series $\sum_{r=1}^{+\infty} \left| C_r(\epsilon) \right|^2 \cos{(r T L/ \hbar)} $ in Eq.(\ref{eq:0}), which are contributed from the bifurcating periodic orbit with period of 4 and its $r$- repetition, generated at the bifurcation point.  In the following, we analyze the eigenfunctions on the Poincar\'e surface of section and directly identify the source of the oscillation.

Figures 6(a)-(c) show numerical plots of the Husimi function $\varrho_n(\phi , \sin\alpha)$ for the energy eigenstates $n=1,2,3,\cdots$, whose representation for the Birkhoff coordinates is derived in Refs.\cite{Crespi,Backer}. The PSS of the classical dynamical system is also plotted in Fig.6(d). One can observe that the individual eigenstates are condensed onto various classical trajectories, which are connected by the quantum tunneling effect. As the system approaches to the semiclassical limit, because of the suppression of quantum tunneling, the energy eigenstates are expected to be localized on one of the classically disconnected regions in the phase space explored by a typical trajectory, such as torus or chaos, and the focus of this article is the localization to the second chaotic region. This property can be analyzed by the integrated Husimi function $I_n$ over the second chaotic region\cite{Makino2000, Makino2018}, 
\begin{equation}
I_n=\frac{ \int_0^1 d\phi \int_{-1}^1 ds \varrho_n (\phi , s) \chi(\phi,s)}{\int_0^1 d\phi \int_{-1}^1 ds \varrho_n (\phi , s)},
\label{eq:XX}
\end{equation}
where $\chi(\phi,s)$ is the characteristic function defined as $\chi(\phi,s)=1$ if $(\phi,s)\in$ 2nd chaotic region and 0 otherwise. The quantity (\ref{eq:XX}) for each eigenfunction provides the degree of localization to a limited area in the phase space. 

Figure 7 shows numerical plots of $I_n$ versus the eigenenergy $\epsilon_n$ at the bifurcation point $\delta_c$. One can observe one large group around $I_n=1$, which consists of eigenfunctions condensing over the second chaotic region, and another large group around $I_n=0$, which consists of eigenfunctions condensing elsewhere. In addition, in the region $0<I_n<1$, there are a lot of eigenfunctions having amplitudes in both the second chaotic region and other regions, which exhibit the quantum tunneling effect. Note that the number of the eigenstates contained in the region $0<I_n<1$ decreases slowly as $\epsilon_n\to+\infty$. Here we classify each eigenstate into the 
group G by the condition $I_n>0.5$ and $\overline{\mbox{G}}$ by $I_n\leq0.5$. The group G is expected to provide the energy level component 
supplied from the eigenstate localized on the second chaotic region in the high energy(semiclassical) limit.

Figures 8(a)-(c) show numerical plots of the TPCF for the three different energy ranges corresponding to Figs.5(a)-(c), respectively. In each figure, the red plots $\rho^{\mbox{\tiny G}} R_2^{\mbox{\tiny G}}(\rho^{\mbox{\tiny G}} L)$ and blue plots $\rho^{\overline{\mbox{\tiny G}}}R_2^{\overline{\mbox{\tiny G}}}(\rho^{\overline{\mbox{\tiny G}}}L)$ represent the normalized TPCF of levels obeying the groups G and $\overline{\mbox{G}}$, respectively, where $\rho^{\mbox{\tiny G}}$ and $\rho^{\overline{\mbox{\tiny G}}}$ are determined by the relative numbers of levels obeying the groups G and $\overline{\mbox{G}}$, and satisfy $\rho^{\mbox{\tiny G}}+\rho^{\overline{\mbox{\tiny G}}}=1$. The black plot represents the TPCF $R_2(L)$ of the whole energy levels. It is quite surprising that, in each figure, the TPCF of the group G(red plots) shows strong spike oscillations whose period is same as the period of the overall TPCF(black plots), while the TPCF of the group $\overline{\mbox{G}}$(blue plots) does not have any spikes at all. From these results, it can be confirmed that the spike oscillation is generated by the eigenstates localized on the second chaotic component in the phase space, where the bifurcation is observed. 

Although the bifurcation we focused on in this paper is a special type, the spike oscillation of the TPCF is possibly observed also in more general bifurcations.  Berry, Keating and Prado investigated the LNV $\Sigma^2(L)$ of the perturbed cat map at the saddle-node  bifurcation and reported an anomalous increase of this quantity at $L\sim p$, which was called the {\it lift-off} effect\cite{BKP}.  This effect has also been observed by Guti\'errez et al. for the pitchfork bifurcation of the coupled quartic oscillators\cite{Marta}.  Since the TPCF is related to the LNV as $\Sigma^2(L) = L+2 \int_0^L (L-x) (R_2(x)-1)dx$ and each spike $R_2(x)>1$ gives a positive contribution to the integral, the {\it lift-off} effect can be expected to arise not only at $L\sim p$ but also at $L\sim 2p, 3p, 4p\cdots$ where the spike oscillations are observed.  Note that $\Sigma^2(L) \sim L + O(L^2) \left[ R_2(0) -1 \right]$ for $L \ll 1$, the spike oscillation at $L=0$ provides little contribution to the {\it lift-off}, which is of the order of $O(L^2)$.  Figure 9 shows numerical plots of the LNV at the bifurcation point $\delta_c$ in our model.  The {\it lift-off} effect is indeed observed at all spike positions $L=p,2p,3p,\cdots$ except $L=0$.  Since the interval $p = 4\pi\sqrt{\epsilon}/l \to+\infty$ as $\epsilon\to+\infty$, the LNV in the semiclassical limit does not show the {\it lift-off} in a finite region.

Makino, Harayama and Aizawa numerically studied the NNLSD of the oval billiard, which characterizes the short-range spectral correlation,  and reported the accumulation between adjacent levels at the bifurcation point\cite{Makino1999}. Since the TPCF is related to the NNLSD $P(0,L)$ as $R_2(L) =\sum_{k=0}^{+\infty}P(k;L)$, where $P(k;L)$ denotes the probability density of finding two levels of spacing $L$ containing $k$ levels in between and $P(k,L)$ is mainly related to the spectral fluctuation characteristic around $L=k+1$, the accumulation $P(0,L)>0$ at $L=0$ corresponds to a spike of the TPCF at $L=0$, which is indeed contributed from the bifurcating orbit.  Let us check the NNLSD at the bifurcation point $\delta_c$.  Figures 10(a)-(c) show numerical plots of the NNLSD for (a) whole energy levels, (b) group G and (c) group $\overline{\mbox{G}}$.  The NNLSD of the whole energy levels shows the accumulation $P(0,L)>0$ at $L=0$ as was reported in Ref.\cite{Makino1999}, and it is apparently contributed from the group G whose NNLSD $P^{\mbox{\tiny G}}(0,L) $ shows 
anomalous increase at $L=0$.

The NNLSD is studied also for the saddle-node bifurcations.  Makino investigated $\rho^{\mbox{\tiny qm}}$ of lemon billiards, which is obtained by applying the Berry-Robnik distribution to the numerical dat of the NNLSD, and observed a sharp decrease of this quantity at $\delta_c$, which is reflecting the accumulation between adjacent levels\cite{Makino2018,Robnik1984}.  The phase space structure of the lemon billiard is partially identical to the oval billiard, some of these bifurcations can be observed also in our model.  One example is the bifurcation at $\delta_{c'}=(16-3\sqrt{23})/7 \sim 0.2304$ ($\eta=1-\delta_{c'} \sim0.77$ for the lemon billiard of Ref.\cite{Makino2018}) where a pair of the elliptic and hyperbolic periodic orbits with period 6 merge as shown in Figures 11(a)-(c).  Note that $\rho^{\mbox{\tiny qm}}(\delta)$ at $\delta_{c'}$ also shows a sharp drop as shown in Fig.4, and the corresponding TPCF certainly shows the spike oscillation as shown in Figure.11(d).

Those works on the LNV and the NNLSD mentioned above imply that the quantum mechanical effect of bifurcation revealed in this paper is more universal and possibly observed also in the transcritical and period-doubling bifurcations in addition to the saddle-node and pitchfork types.   We need to gather more cases in the future works.   

\section{Conclusion and Discussion}
In summary, the two-point correlation function (TPCF) of the quantum oval billiard whose classical dynamical system shows the bifurcation, was investigated by the numerical experiments. The sequence of energy levels at the bifurcation point showed a remarkable accumulation at a specific energy interval, which was observed as a periodic spike oscillation of the TPCF. To identify the source of this oscillation, we analyzed the Husimi representation of energy eigenfunctions which in the semiclassical limit is expected to localize onto one of the phase space components, such as torus and chaos, and divided individual energy levels into the spectral components. We have found that the periodic spike oscillation is contributed from the eigenfunctions that is localized onto the second chaotic region where the bifurcation is observed. We also analyzed the TPCF in terms of the Gutzwiller periodic orbit theory and found that the period of the spike oscillation agrees well with the fundamental period of the component of the semiclassical TPCF, which is contributed from the hyperbolic periodic orbit with period 4 generated at the bifurcation point. These results enable us to conclude that the bifurcation of the classical dynamical system generates the strong spectral accumulation in the corresponding quantum system that is observed as the periodic spike oscillations of the TPCF.  

The spectral accumulation is one of the interesting features in the research field of level statistics that may provide useful knowledge for the quantum computation and the quantum annealing closely related to the non-adiabatic transition\cite{Nielsen}, and one well-known example 
arises from the time reversibility or the spatial reversibility\cite{Shn,Chirikov,Frahm, CK, Makino2003, Makino2009}.  This type of accumulation is quite similar to our result shown in Sect.3, causing spike oscillations of NNLSD and TPCF.  However, its mechanism is different from our result arising from the bifurcation.  Another example is caused by the localization of eigenfunctions.  Molchanov investigated a one-dimensional Schr\"odinger operator with a random potential and showed the spectral accumulation arose from the localization of eigenfunctions in the semiclassical limit\cite{Mol}. Minami also studied a one-dimensional Schr\"odinger operator with $\delta$ potentials and reported a similar result\cite{Minami}. The spectral accumulation observed in Sect.3 is also related to the localization of eigenfunctions around the bifurcating periodic orbits. 

The present paper focused exactly on the bifurcation point and observed the anomalous accumulation of levels, while one may expect from the plot of Fig.4 to observe this phenomenon also in the vicinity of the bifurcation point, so that it is also interesting to determine the parameter region in terms of the periodic orbit theory, where the quantum mechanical effect of bifurcations is observed.  The range of this region will be determined by evaluating the amplitude $C_{j^*}(\delta)$ of the bifurcating periodic orbit $j^*$.  This part will be investigated elsewhere.

\section*{Acknowledgment}
This work is supported by JSPS KAKENHI No.15K13538 to H.M.

\appendix
\newpage
\begin{figure}[ttbp]
\begin{center}
\includegraphics{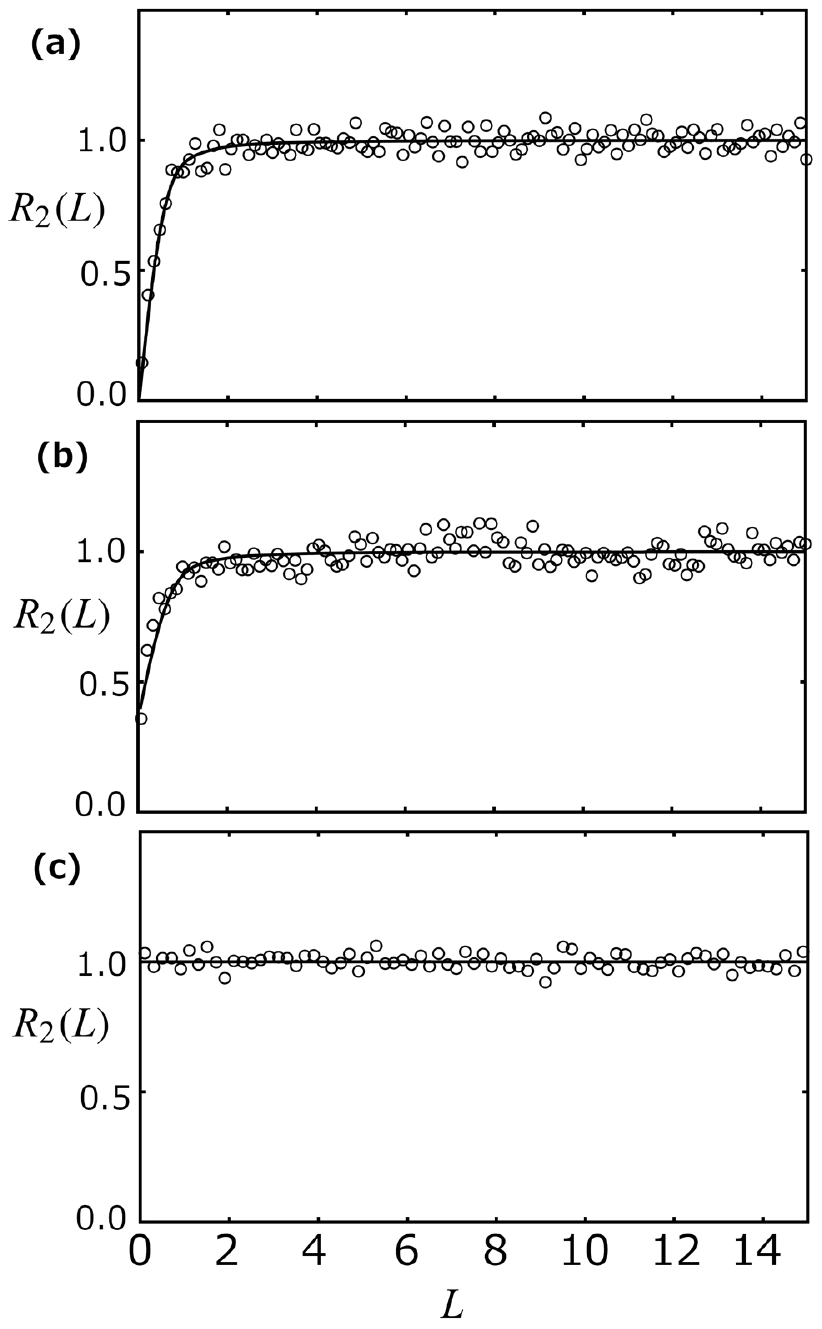}
\end{center}
\caption{Numerical plots of the TPCF $R_2(L)$ for (a)$\delta=0$, (b)$\delta=0.39$, and (c) $\delta=1$, which are computed by 6000 levels starting from $\epsilon=8000$. The solid line represents the formula $R_2(\rho^{\mbox{\tiny qm}}, L)$ with (a) $\rho^{\mbox{\tiny qm}}=1.0$, (b) $\rho^{\mbox{\tiny qm}}=0.77$(the best fitting value) and $\rho^{\mbox{\tiny qm}}=0.0$. }
\label{fig3}
\end{figure}
\begin{figure}[ttbp]
\begin{center}
\includegraphics{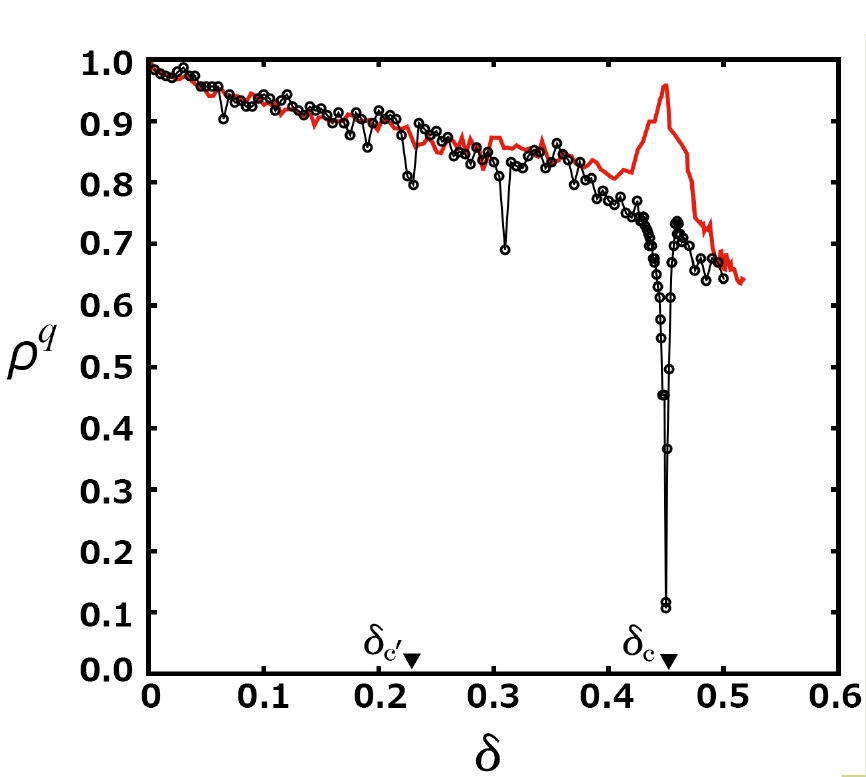}
\end{center}
\caption{Numerical plots of $\rho^{\mbox{\tiny qm}}(\delta)$ computed by the best fitting curve of the formula $R_2(\rho,L)$ to the quantum mechanical data which is obtained by 6000 levels from $\epsilon=8000$. The red curve represents the relative phase volume(Liouvill\'e measure) of the chaotic region in the classical dynamical system.}
\label{fig4}
\end{figure}
\begin{figure}[ttbp]
\begin{center}
\includegraphics{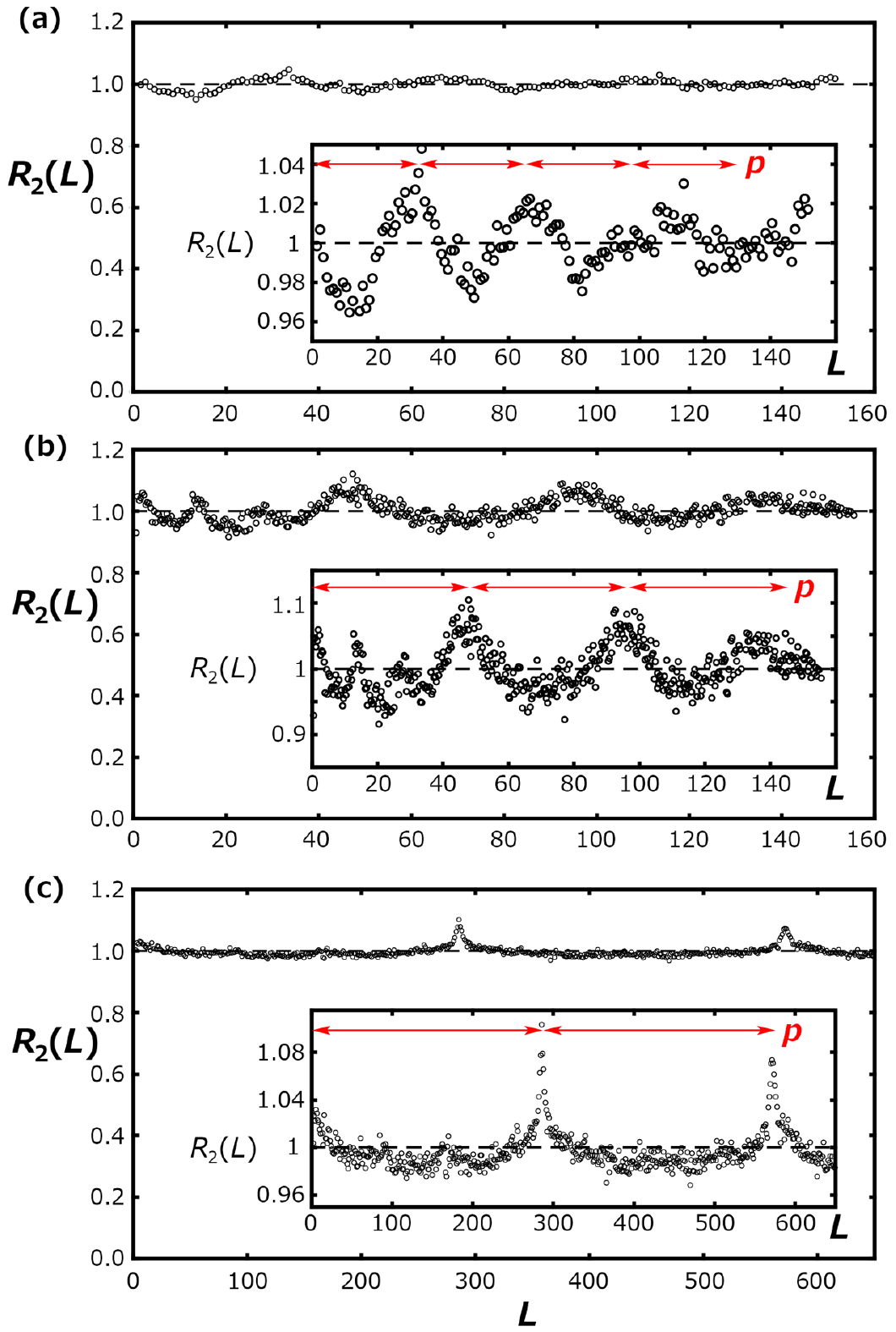}
\end{center}
\caption{Numerical plots of the two-point correlation function $R_2(L)$ at the bifurcation point $\delta_c$, computed by (a) 6000 levels from $\epsilon=2000$, (b)6000 levels from $\epsilon=8000$ and (c) 12000 levels from $\epsilon= 4 \times 10^5$. The inset in each figure shows the enlargement around $R_2(L)=1$.  $p=4\pi\sqrt{\epsilon}/l\sim0.46\sqrt{\epsilon}$ is the fundamental period of the component in equation (2) related to the bifurcating periodic orbit, and this value is calculated by the medium value of the energy range.}
\label{fig5}
\end{figure}
\begin{figure}[ttbp]
\begin{center}
\includegraphics{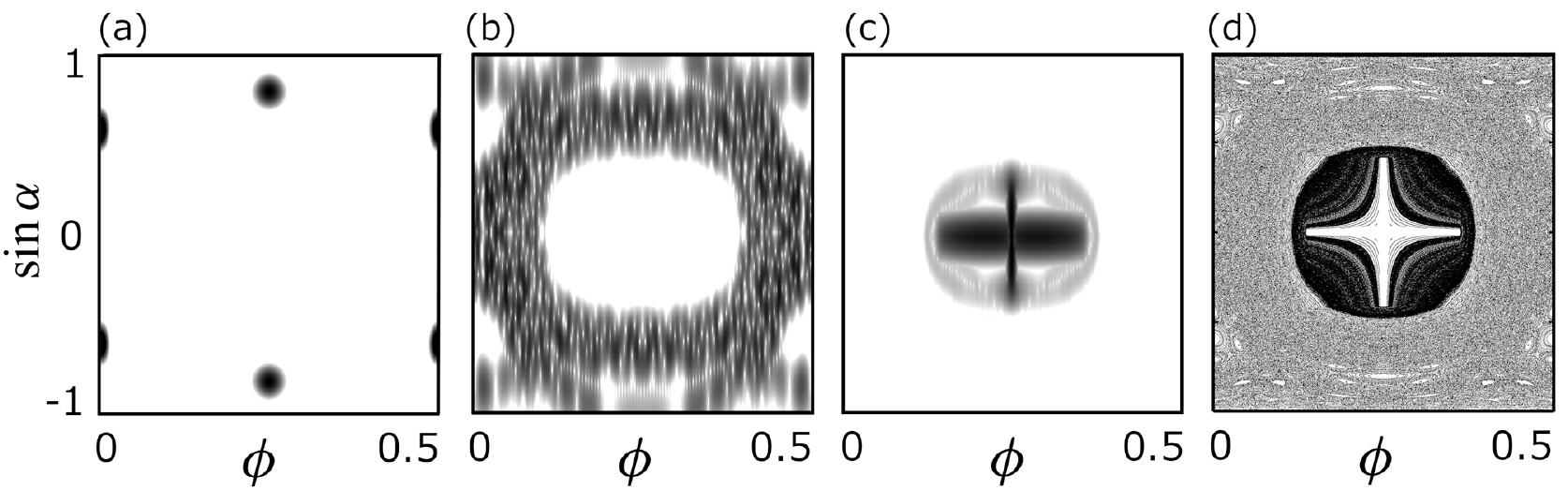}
\end{center}
\caption{Contour density plots of the Husimi function $\varrho_n(\phi , \sin\alpha)$ for typical eigenstates at $\delta=\delta_c$:(a) $\epsilon_n= 
404108.62$, (b)$\epsilon_n= 404280.89$, (c)$\epsilon_n= 404236.36$, (d) Poincar\'e surface of section of the classical dynamical system. .}
\label{fig6}
\end{figure}
\begin{figure}[ttbp]
\begin{center}
\includegraphics{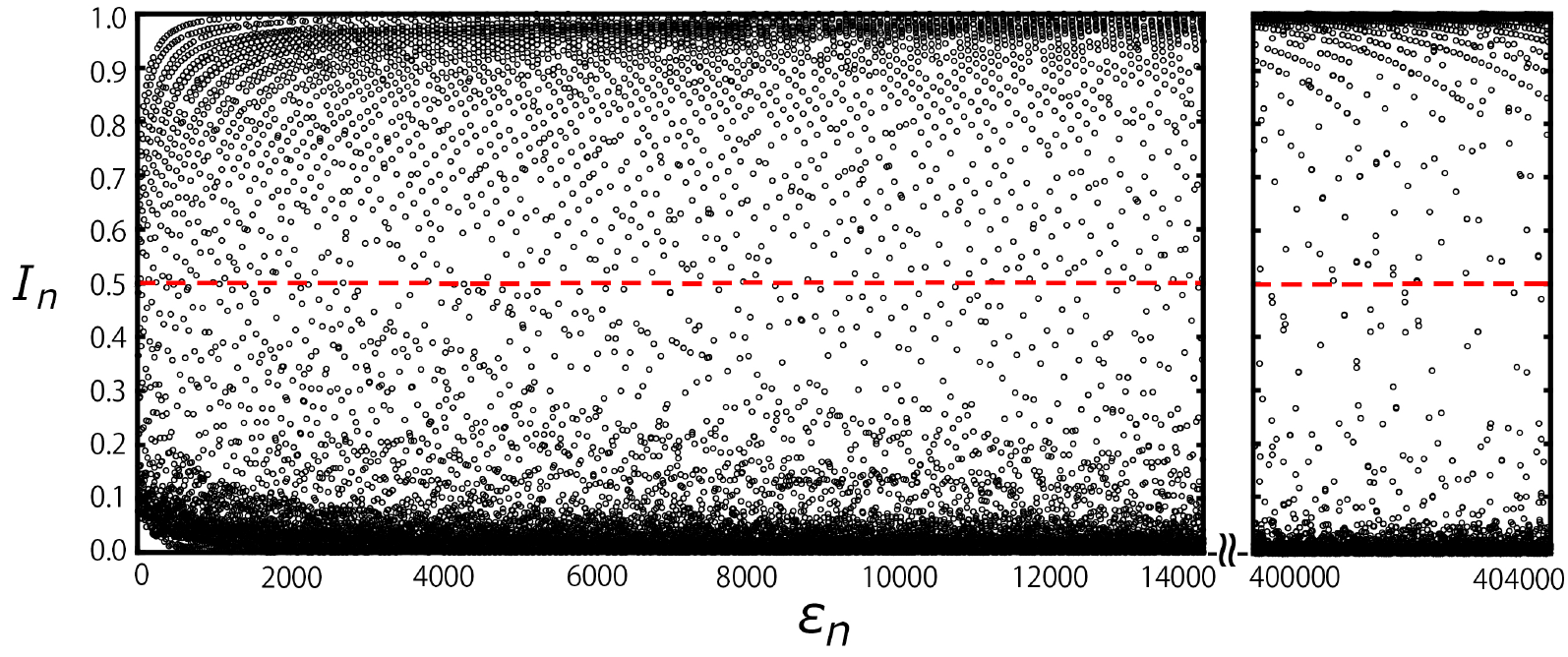}
\end{center}
\caption{Numerical plots of the quantity $I_n$ versus the eigenenergy $\epsilon_n$ for the quantum eigenstates at the bifurcation point $\delta_c$. The red line 
is the threshold which approximately divides all eigenstates $n=1,2,3\cdots$ into two groups G($I_n>0.5$) and $\overline{\mbox{G}}$($I_n\leq 0.5$). }
\label{fig7}
\end{figure}
\begin{figure}[ttbp]
\begin{center}
\includegraphics{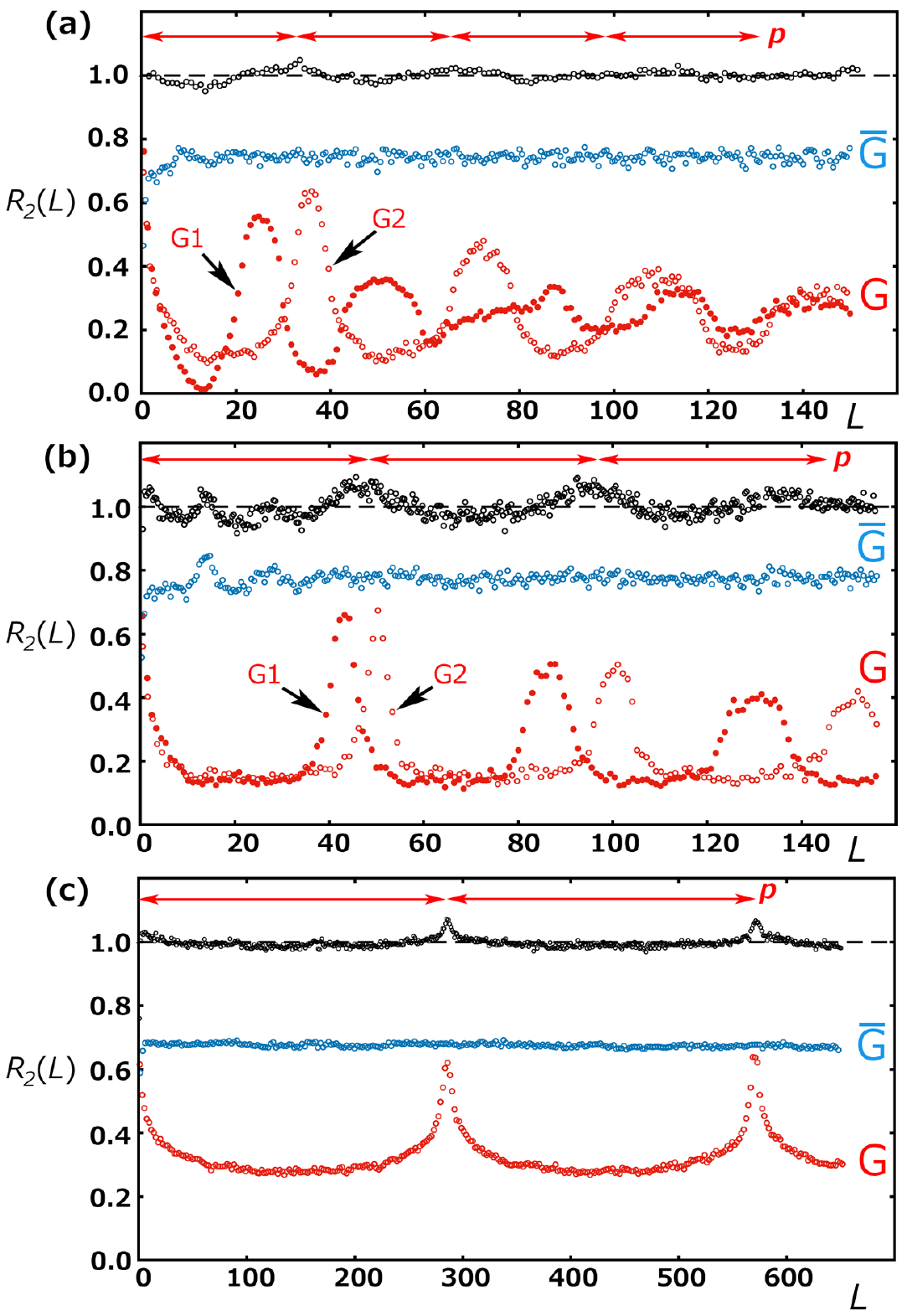}
\end{center}
\caption{Numerical plots of the TPCF at the bifurcation point $\delta_c$, which are obtained for the three energy ranges corresponding to Figs.5(a)-(c); (a) 6000 levels from $\epsilon=2000$; (b)6000 levels from $\epsilon=8000$; (c) 12000 levels from $\epsilon=4\times 10^5$. The black plot represents the TPCF $R_2(L)$ of the whole energy levels, and the red plots $\rho^{\mbox{\tiny G}} R_2^{\mbox{\tiny G}}(\rho^{\mbox{\tiny G}} L)$ and blue plots $\rho^{\overline{\mbox{\tiny G}}}R_2^{\overline{\mbox{\tiny G}}}(\rho^{\overline{\mbox{\tiny G}}}L)$ represent the normalized TPCF of levels obeying the groups G and $\overline{\mbox{G}}$, respectively. To reduce the range of the period $p=4\pi\sqrt{\epsilon}/l$ in the lower energy regions, the group G is further divided into the two sub-groups G1 and G2 which are extracted respectively from narrower ranges $\epsilon_n\in[2000,5000]$ and $\epsilon_n \in[5000,8000]$ in Fig.8(a), and extracted respectively from $\epsilon_n\in[8000,11000]$ and $\epsilon_n\in[11000,14000]$ in Fig.8(b).}
\label{fig8}
\end{figure}
\begin{figure}[ttbp]
\begin{center}
\includegraphics{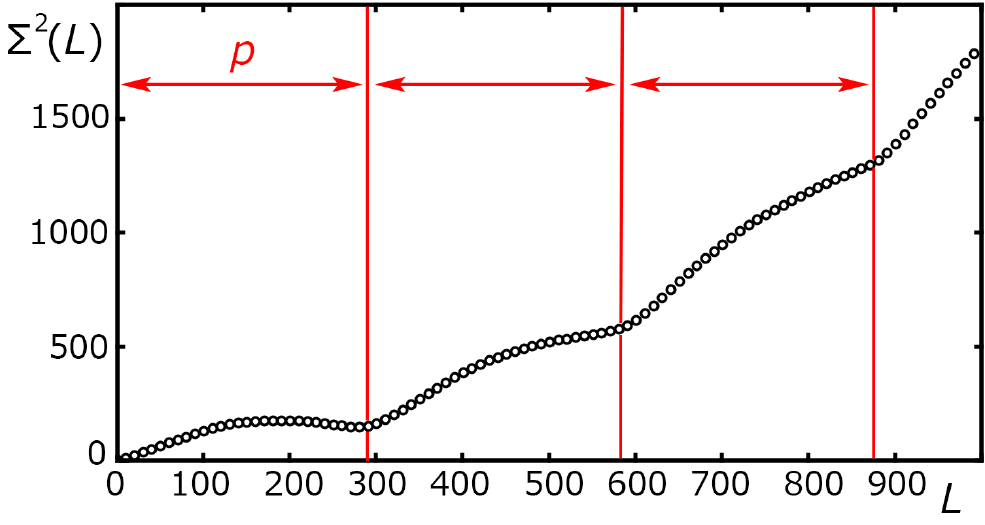}
\end{center}
\caption{Numerical plots of the level number variance(LNV) $\Sigma^2(L)$ at the bifurcation point $\delta_c$, computed by using 12000 levels from $\epsilon=4\times 10^5$.}
\label{fig9}
\end{figure}
\begin{figure}[ttbp]
\begin{center}
\includegraphics{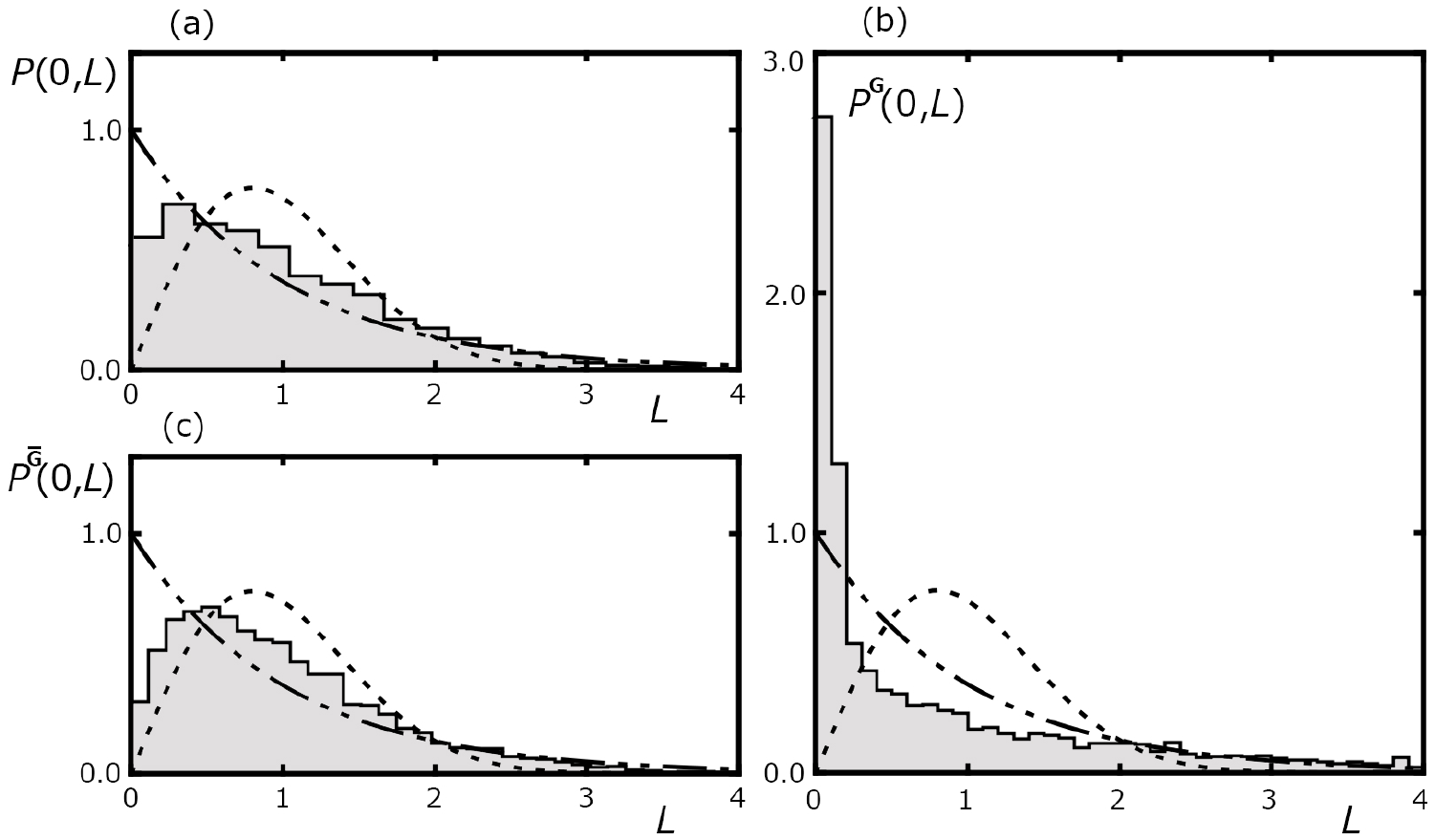}
\end{center}
\caption{Numerical plots of the nearest-neighbor level-spacing distribution(NNLSD) at the bifurcation point $\delta_c$, obtained by (a) the whole energy levels, (b)energy levels obeying the group G, (c) energy levels obeying the group $\overline{\mbox{G}}$.  We have used 12000 levels starting from $\epsilon=4\times 10^5$. The dashed and double-dashed lines represent the NNLSD resulting from the GOE statistics and the Poisson statistics, respectively. }
\label{fig10}
\end{figure}
\begin{figure}[ttbp]
\begin{center}
\includegraphics{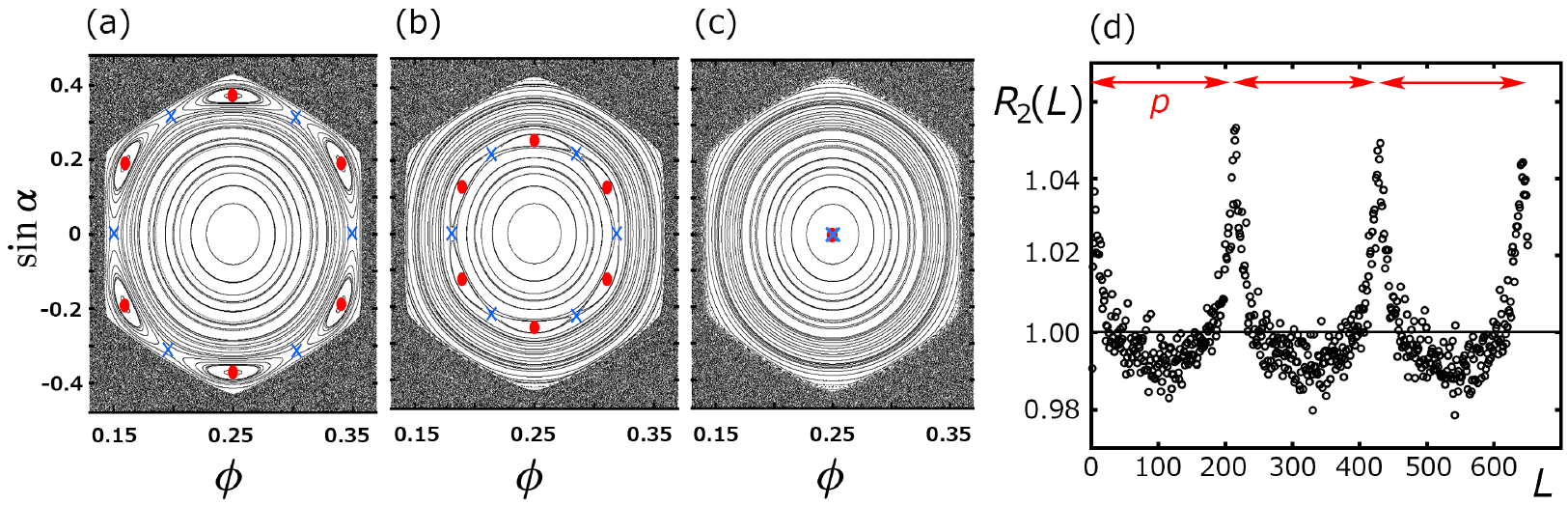}
\end{center}

\caption{Poincar\'e surface of section for (a) $\delta=0.2263$, (b)$\delta=0.2285$, (c) $\delta=0.2304\sim \delta_{c'}$ 
where the red and blue marks correspond to the elliptic and hyperbolic points with period 6, respectively; (d) Two-point correlation function $R_2(L)$ at the bifurcation point $\delta_{c'}$, computed by 16000 levels from $\epsilon=4\times 10^5$.  The period of the spike oscillation is in good agreement with the fundamental period 
$p\sim 0.34\sqrt{\epsilon}$ of the periodic orbit sum in the trace formula (\ref{eq:0}), which is contributed from the bifurcating periodic orbits.}
\label{fig11}
\end{figure}

\end{document}